\documentclass[sigconf]{acmart}
\copyrightyear{2024}
\acmYear{2024}
\setcopyright{rightsretained}

\copyrightyear{2024}
\acmYear{2024}
\setcopyright{acmlicensed}\acmConference[ICMI '24]{INTERNATIONAL CONFERENCE ON MULTIMODAL INTERACTION}{November 4--8, 2024}{San Jose, Costa Rica}
\acmBooktitle{INTERNATIONAL CONFERENCE ON MULTIMODAL INTERACTION (ICMI '24), November 4--8, 2024, San Jose, Costa Rica}
\acmDOI{10.1145/3678957.3678960}
\acmISBN{979-8-4007-0462-8/24/11}

\AtBeginDocument{%
  }

 \acmConference{ICMI ’24} {4-8 November 2024}{ Costa Rica}

\begin{document}

\title{RealSeal: Revolutionizing Media Authentication with Real-Time Realism Scoring}

\author{Bhaktipriya Radharapu} 
\email{}

\author{Harish Krishna}
\email{}



\begin{abstract}

The growing threat of deepfakes and manipulated media necessitates a radical rethinking of media authentication. Existing methods for watermarking synthetic data fall short, as they can be easily removed or altered, and current deepfake detection algorithms do not achieve perfect accuracy. Provenance techniques, which rely on metadata to verify content origin, fail to address the fundamental problem of staged or fake media.

This paper introduces a groundbreaking paradigm shift in media authentication by advocating for the watermarking of real content at its source, as opposed to watermarking synthetic data. Our innovative approach employs multisensory inputs and machine learning to assess the realism of content in real-time and across different contexts. We propose embedding a robust realism score within the image metadata, fundamentally transforming how images are trusted and circulated. By combining established principles of human reasoning about reality, rooted in firmware and hardware security, with the sophisticated reasoning capabilities of contemporary machine learning systems, we develop a holistic approach that analyzes information from multiple perspectives.

This ambitious, blue sky approach represents a significant leap forward in the field, pushing the boundaries of media authenticity and trust. By embracing cutting-edge advancements in technology and interdisciplinary research, we aim to establish a new standard for verifying the authenticity of digital media.
\end{abstract}


\maketitle
\section{Introduction}
Manipulated images facilitated by accessible editing software, have become pervasive threats, particularly within the political sphere. Despite the implementation of robust safeguards, popular generative AI systems can still be attacked \cite{10413929} to produce harmful and photorealistic images that violate their terms of service \cite{dalle, googleGenerativeProhibited}. Lying through images is not a new phenomenon; historical examples, such as the infamous doctored photo of Stalin \cite{historyPhotosBecame} where political enemies were airbrushed out, and more recent instances like the deepfake images of Donald Trump with Black voters \cite{bbcTrumpSupporters} illustrate how image manipulation has long been used to misrepresent reality. Deepfakes, which involve AI-generated synthetic media, add a new layer of complexity to this issue. The ease of using tools such as Stable Diffusion \cite{podell2023sdxl} had recently led to Twitter being inundated with seemingly photorealistic images of Taylor Swift in compromising positions \cite{cbsnewsBlocksSearches}. These manipulations, whether through traditional photoshopping or AI deepfake technology, undermine trust in media, erode the credibility of public figures and business leaders, and threaten democratic and economic processes by fabricating events, discrediting opponents, and inciting social discord. 

There are a few ways this problem has been tried to be solved \cite{pei2024deepfake, wang2022deepfake}. 
One way is to use deep learning or AI based methods to discern between real and modified images. This might involve automating the search for visual inconsistencies like an incorrect number of fingers or the curvature of what must be a straight line. However, recent studies  \cite{tariq2021i, beckmann2023fooling} have shown that current detection methods struggle to effectively identify deepfakes, highlighting the need for more advanced and robust solutions. Moreover, the image generation methods are constantly evolving at a rapid pace, and are often trained to fool models that can detect if an image was made by AI. Such 'reactive' methods identify deepfakes only after creation, and may not be effective against the most convincing fakes, or be credible to establish authenticity.

An alternative approach to addressing the issue of manipulated images is to automatically embed machine generated watermarks in all AI-generated or edited content, similar to the methods employed by image generation models like Deepmind's Imagen \cite{deepmindSynthID}, Adobe Firefly \cite{thevergeAdobeCreated}, and DALL-E 3 \cite{thevergeOpenAIAdding}.
However, these watermarks can be removed or lost when images are converted, compressed, or uploaded to websites that strip metadata \cite{hackerfactorProblemsWith}. Moreover these watermarks are while robust to some modifications, can still be fooled by imperceptible pixel changes
\cite{wiredResearchersTested, fortuneMetasC2PA}, rendering them ineffective.
Moreover, there are always going to be editing and generation tools that do not use these watermarks. Additionally, fake watermarks can be created to deceive detectors. 

As more and more AI image manipulation techniques become ubiquitous, this way of trying to weed out all the fake ones does not work. Instead, one way to look at the problem would be to affix a certificate of authenticity with only the real images and hope that fake images can't acquire one.

C2PA \cite{c2paOverviewC2PA} is a hardware-based approach based on this idea that integrates cryptography to embed provenance information as metadata within media files. It suffers from a number of issues \cite{hackerfactorC2PAFrom, ieeeMetasFlimsy}. C2PA relies on trust in the entity adding the metadata, leaving it vulnerable to manipulation and impersonation \cite{hackerfactorC2PAFrom}. Metadata alone doesn't verify the content itself, leaving room for misleading information \cite{fooling_adobe}. Hardware watermarks cannot prevent the capture and manipulation of fake content from screens or recordings \cite{wiredWontStop}. Camera hardware can be altered, and post-processing is common in social media and news outlets. Only signed providers or owners of specialised hardware on the network can add a seal of provenance. Not only does this have democratisation issues, but there is nothing stopping trusted people themselves to endorse fake content as it happened with BBC's "verification" of a forged video \cite{hackerfactorIEEEBBC}.

In this work, we address these limitations by drawing inspiration from human perception \cite{scientificamericanMakingSense, nihPerceptionSynchrony}, which excels at understanding broader contexts through contextual awareness, multisensory input, intuition, and environmental interaction. Current imaging systems, in contrast, capture a limited perspective, relying mainly on visual and auditory information, making them susceptible to manipulation and lacking interactive capacity. To achieve human-level reasoning, imaging systems need enhancements such as multisensory input, spatial reasoning, and contextual understanding, allowing them to gather and process information similarly to human perception.

Our proposed solution builds on technologies like FaceID \cite{faceid} or FaceUnlock \cite{googleUnlockYour}, which can reliably distinguish real faces from manipulated images or masks on a smaller scale \cite{wiredSecureIPhone}. By extending this concept to a larger scope, we envision autonomous systems capable of accurately assessing and understanding real-world events through human-like perception. What sets our idea apart is its unexplored territory. While current technologies focus on detecting manipulated media after its creation, our approach aims to prevent the creation of deepfakes by embedding realism scores directly into images.

Furthermore, advancements in secure firmware and hardware solutions \cite{nistNISTSpecial}, seen in iOS  \cite{apple_secure_enclave} and Android \cite{wiredGooglesFeature} devices, provide a robust foundation for preventing tampering at the operating system level. If tampering is detected, our system can employ countermeasures such as data deletion or blacklisting, reducing the risk of compromised reality scores.

Implementing this technology could lead to new universal standards for digital media authenticity, fostering greater trust across platforms and industries. This could be particularly transformative for sectors that heavily rely on the veracity of digital content.

\section{Our Method}

\begin{figure*}
\centering
\includegraphics[width=\textwidth]{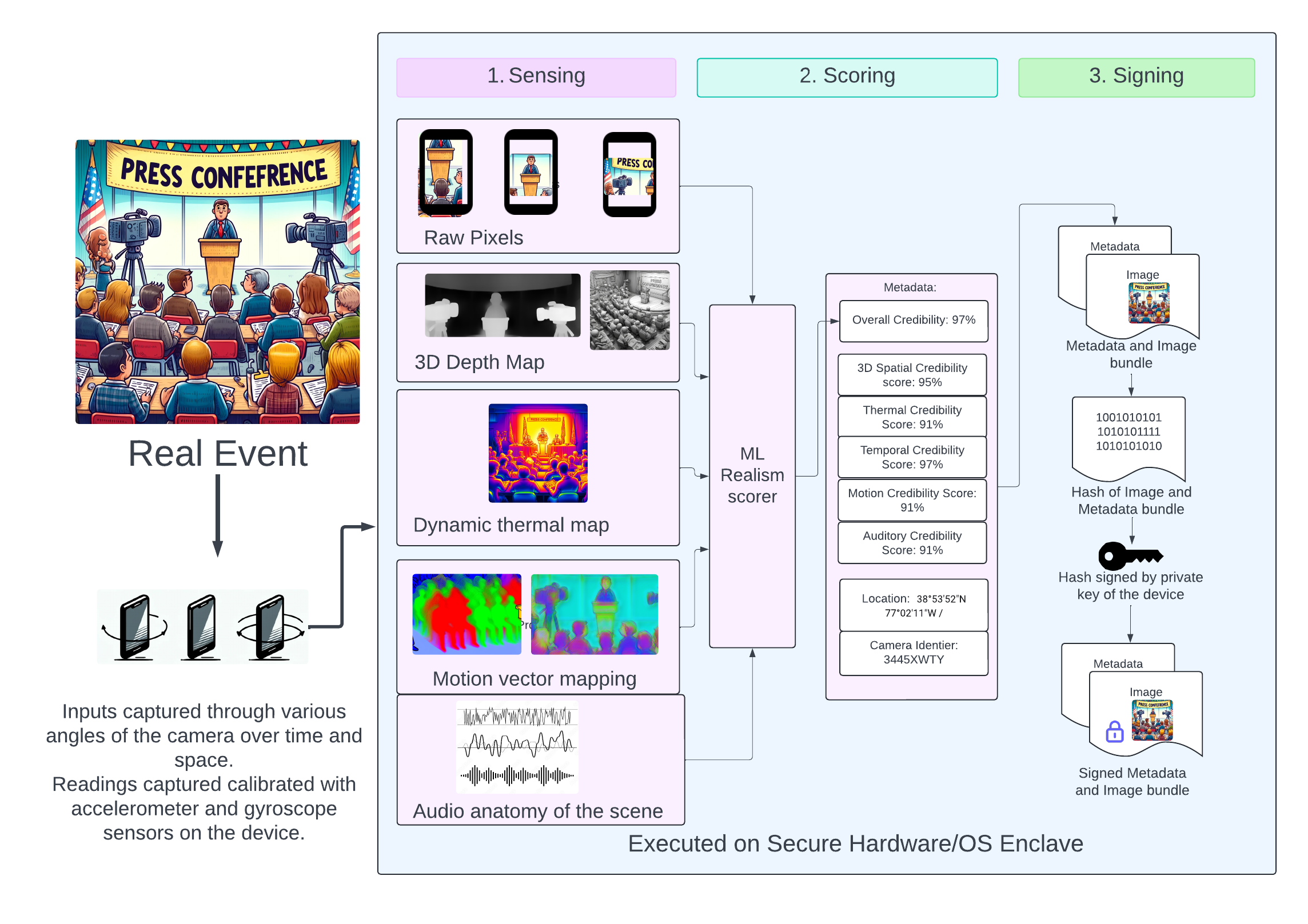}
\caption{Overview of the RealSeal approach. The source-based watermarking approach inherently enhances the reliability and security of digital media authentication. }
\label{fig:my_image}
\end{figure*}

Our method draws inspiration from human perception and reasoning about the reality of events. Humans use visual, auditory, and temporal signals, and observe the motion of objects over space and time to assess the authenticity of events. Mimicking this process, our approach involves three key steps: sensing, scoring, and signing.

First, we capture data in real-time using a range of sensors. This step, termed \textbf{Sensing}, involves collecting diverse data points about the event. Next, in the \textbf{Scoring} step, these data points are processed through a machine learning model to generate a realism or credibility score for the event. Finally, in the \textbf{Signing} step, we cryptographically sign \cite{wikipediaDigitalSignature} the image with the realism score embedded in the metadata, ensuring the integrity and authenticity of the data.

The entire pipeline, from data capture to processing and watermarking, operates within a secure environment at the OS level to prevent tampering. Users can verify the image's authenticity using the device's public key and viewing the realism score in the metadata. This process ensures that any edits to the image will invalidate the signature, maintaining the integrity of the captured event.

\subsection{Multimodal Sensing:} 

\paragraph{Reasoning about the space visually:} 
Traditional imaging systems often only provide two-dimensional (2D) representations, which can severely limit the accuracy of scene perception. Relying solely on focal length for estimating depth is not only inadequate but also vulnerable to manipulation. For example, it is feasible to produce C2PA-vetted images using counterfeit lenses, as documented by HackerFactor \cite{hackerfactorC2PAapossButterfly}. This technique can deceitfully create an illusion of depth in screen recordings viewed through a C2PA-compliant camera. This underscores the critical need for full 3D reasoning capabilities that can more effectively discern genuine content from artificially staged scenes.

To address this, we propose integrating infrared (IR) depth sensors to create depth maps, thereby capturing and understanding scenes in three dimensions (3D). This technology, already used on a small scale in Face ID \cite{true_depth_camera} systems to prevent spoofing, can either flood the scene with IR beams or use LIDAR to map the environment in 3D, similar to its application in AR/VR rendering on smartphones \cite{researchUDepthRealtime}. By capturing rich 3D depth data, we can better distinguish between real and fake scenes on a larger scale, such as determining if a politician speaking at a conference is genuine or if it is a deepfake played on a high-definition screen.
\paragraph{Audio data}: 
We also capture audio cues through a microphone over time, adding an auditory dimension to the system's perception capabilities. By analyzing these audio cues, such as ambient sounds and lip movements, speech patterns, the system can detect anomalies in audio-visual synchronization as shown in recent studies \cite{Yang2023AVoiDDFAJ, Haliassos2022LeveragingRT}. These discrepancies can be indicative of manipulated content.

\paragraph{Temporal Motion data}: Additionally, we recommend capturing audio and visual inputs from multiple angles by moving the camera and tracking changes in depth maps and pixel values over time. Research \cite{zhang2021consistent, gkioxari2019mesh} has demonstrated that machine learning models can better understand the world through temporal sequences, depth information, pixel values, and by capturing disparities and bridging 2D and 3D data across different camera angles and motion.

Using motion sensors like accelerometers and gyroscopes to monitor camera orientation further enhances spatial reasoning. This method mirrors human perception, where moving around and observing a scene from different perspectives provides a more accurate assessment than a single, static shot.

We also incorporate temporal reasoning by taking multiple shots of a scene over time as done in past research to capture inconsistencies across frames \cite{gu2022region,gu2022hierarchical, zhang2024spatiotemporal}. By capturing optical flow, which tracks how objects move within the scene, we can judge alignment between depth maps, pixel values and motion patterns \cite{10445863, roxas2018real}. We employ a multi-level spatial-temporal feature aggregation and alignment-based approach to assess the consistency between various inputs and determine if a scene is real or staged. This method makes it challenging to convincingly fake an event across multiple frames. 

\paragraph{Thermal data}:We propose incorporating thermal imaging capabilities into the system to capture temperature maps of objects within the scene, adding a thermal dimension to our analysis. This provides additional context for assessing authenticity. Research indicates that machine learning systems can leverage 3D understanding and temperature data to develop realistic priors for thermograms \cite{rangel20143d, doi:10.1080/17686733.2021.1991746}. For example, human body temperature typically averages around 98.6°F (37°C), but this can vary across different body parts. If thermal imaging captures a screen showing a politician speaking with a uniform temperature of 37°C, it may indicate something unusual, suggesting the content was recorded in a controlled environment rather than a natural setting, pointing to potential manipulation.

For instance, objects exposed to natural sunlight exhibit a temperature gradient, whereas artificial scenes often lack this variability. Similarly, an engine running in a vehicle would display varying heat signatures, unlike a static prop. These subtle differences in temperature distribution help the system identify and flag content that may have been staged or digitally altered.

\subsection{Credibility Scoring and Reasoning:}
Our method involves comprehensive Reality Score Modeling and Reasoning based on the multi-dimensional inputs captured by the system. We compute individual credibility scores across various dimensions—3D spatial, thermal, motion, and auditory—before aggregating these into an overall credibility score. This process ensures a holistic evaluation of the scene's authenticity.
\subsection{Signing}

Once individual credibility scores for each dimension are determined, the model computes the overall credibility score of the event. This overall score, along with additional metadata such as the date, camera ID, location, and individual dimension scores, forms the metadata of the image. This metadata is bundled with the image, creating a data package.

We then sign this data package using standard cryptographic signing schemes \cite{wikipediaDigitalSignature, cryptographyx2014Cryptography}. A cryptographic hash function \cite{md5,wikipediaSHA2Wikipedia} is applied to the image and metadata bundle, generating a unique fixed-length hash value that acts as a digital fingerprint. Any alteration to the content, even by a single pixel, or any change to the metadata, such as tampering with the realism score, will produce a different hash value. The hash value is encrypted using a private key specific to the authorized device, creating a digital signature.

This digital signature can be verified using the corresponding public key, which is publicly available (for example, from the device manufacturer's website, where private and public key pairs are generated during manufacturing for untampered devices). When the image and metadata bundle are shared, the recipient can verify its authenticity by hashing the bundle on their end and comparing the hash with the decrypted signature (obtained using the public key of the device).

By implementing cryptographic signing, we ensure that the image and metadata bundle remains untampered from the moment of capture. Each digital signature acts as a seal of approval, guaranteeing the authenticity and integrity of the media throughout its distribution. This signing process also allows our approach to be integrated into existing provenance mechanisms, such as blockchain or the Content Credentials Provenance Architecture (C2PA) \cite{c2paOverviewC2PA}, establishing a verifiable trace back to the device that originally captured the image.

\subsection{Secure execution environment}
Preventing device tampering is crucial for ensuring the system's reliability. The entire system relies on having a trusted device\cite{7345265} with untampered and functional sensors, which accurately capture inputs from the real scene rather than from a manipulated stream. Additionally, the machine learning model's weights must remain untampered, with no leaks of the private key and no alteration of metadata before it is signed. Achieving this level of security requires meticulous synchronization between hardware and software components. If any part is compromised, the authenticity scores cannot be trusted.

To achieve this, we propose methods similar to those used by modern operating systems to prevent jailbreaking and tampering \cite{10.1007/3-540-61996-8_49}. One approach is to use a secure bootloader \cite{ocp_secure_boot_spec}, ensuring that the system camera and operating system always boot into a secure, signed version of the OS. If the system detects any tampering, it will not boot. All critical operations, including model inference, key management, sensor data processing, and signing, should occur within the secure version of OS.

Another approach is to utilize Secure Enclave or Trusted Execution Environment (TEE) technologies, which are isolated execution environments within the main processor \cite{intel,amdTechnicalInformation,androidTrustyAndroid}. An enclave is an isolated area of memory with sensitive application data protected by the CPU. The code and data inside an enclave are encrypted and only accessible from within the enclave itself. Enclaves provide a highly secure environment for processing sensitive data, even if the operating system, BIOS, or other system components are compromised. Several trusted environment solutions have been built of secure enclaves \cite{arnautov2016scone, 5416657, brandao2021hardening} .These environments can securely perform sensitive operations such as sensor reading, model inference, key reading, signing, and metadata embedding with the reality score. This isolation protects these processes from unauthorized access or manipulation from the user .

Security is necessary not only for computation but also for storage. We recommend using Hardware Security Modules (HSMs) \cite{Sommerhalder2023, nistFederalInformation}, specialized devices designed to securely store and manage sensitive data like cryptographic private keys and model weights. HSMs protect against unauthorized access, key leakage, and tampering with model weights.

We advocate for firmware-level implementations and tight integration of software, OS, and hardware techniques to ensure tamper-proof operation as done in military grade systems \cite{militaryembeddedDesigningImplementing}. This provides a reliable seal of authenticity, securing the entire pipeline—from capturing sensor inputs to ML model inference and the signing of the image.

\section{Conclusion and Limitations}

The proposed source-based watermarking method offers several advantages over existing techniques. Unlike strong encryption, which does not inherently increase data trustworthiness, our method begins with a reality score assessment, evaluating media authenticity during creation rather than afterward. This provides not only provenance but also credibility of the scene’s realness \cite{hackerfactorIEEEBBC}.

Our system’s numerous guardrails make it significantly harder to fake. Traditional systems suffer from the analog hole vulnerability, where an attacker could photograph a high-quality fake printout with a crypto-signing camera, manipulating settings to match the alleged scene. Metadata can be bypassed with manual lenses or diopter filters \cite{fooling_adobe}. Low-cost attacks might use diffusers, soft lenses, long exposures, or multiple projectors for higher resolution \cite{fooling_adobe}. Our method mitigates such attacks by capturing multisensory input, including depth maps, motion, temperature, and tactile feedback. While individual components can be faked, replicating an entire multisensory environment is challenging.

Additionally, our method secures the signal chain. Unlike other state of the art systems, like Truepic and Qualcomm, where an attacker could tamper or forge the input stream of sensor data \cite{hackerfactorClosedStandards}, we integrate secure hardware and software components to ensure a secure environment, where camera sensors remain uncompromised for a signature to be generated.  Our system prevents firmware tampering with robust hardware security measures.

We acknowledge that even with high reality scores, there may be ways to capture real but misleading or staged scenes, such as using body doubles, masks of famous people, or strategically avoiding showing faces. However, we argue that this is not primarily a cryptographic or technical problem but an issue out of scope for such a system. Our system can aid user discernment by leveraging face detection and recognition technologies to identify individuals in the captured media, though we recognize this issue.

Another limitation is that not all image manipulations are harmful; some are simple crops, or compression. However, these benign changes still alter the image signature, removing the authenticity score, and are thus not certified to be free of modification. 

Our goal is to provide users with comprehensive metadata and reasoning to make informed judgments, treating unmarked content skeptically. By quantifying authenticity with a "reality score" backed by transparent multisensory input and reasoning, we empower critical evaluation even when logical deceptions occur.

In essence, our method's strength lies in combining multisensory capture, secure hardware/software integration, and transparent reasoning, creating a robust foundation inaccessible to conventional manipulation techniques. It raises the bar significantly with respect to image manipulation and aligns with establishing societal trust in our increasingly visual digital landscape.

\bibliographystyle{plain}
\bibliography{sample-base}

\appendix

\end{document}